\documentclass[12pt]{article}
\usepackage{epsf}
\title{ {\bf
Charged lepton electric dipole moments with the localized leptons
and the new Higgs doublet in the two Higgs doublet model}}
\author{\vspace{1cm}\\
        {\bf E. O. Iltan}
        \thanks{E-mail address:
        eiltan@newton.physics.metu.edu.tr}
 \\
        Physics Department, Middle East Technical University \\
        Ankara, Turkey\\}

\date{}

\begin{document}
\setlength{\baselineskip}{24pt}
\maketitle
\setlength{\baselineskip}{7mm}
\begin{abstract}
We study the lepton electric dipole moments in the split fermion
scenario, in the two Higgs doublet model, where the new Higgs
scalars are localized around the origin in the extra dimension,
with the help of the localizer field. We observe that the
numerical value of the electron (muon, tau)  electric dipole
moment is at the order of the magnitude of $10^{-31}\,
(10^{-24},\, 10^{-22})\, (e-cm)$ and this quantity is sensitive
the new Higgs localization in the extra dimension.
\end{abstract}
\thispagestyle{empty}
\newpage
\setcounter{page}{1}
\section{Introduction}
The electric dipole moments (EDMs) of fermions are worthwhile to
study since they are driven by the CP violating interactions. CP
violation is carried by the complex Cabibo Kobayashi Maskawa (CKM)
matrix elements in the quark sector and the possible lepton mixing
matrix elements in the lepton sector, in the framework of the
standard model (SM). The negligibly small theoretical values of
the fermion EDMs in the SM forces one to search the physics
beyond, such as multi Higgs doublet models (MHDM), supersymmetric
model (SUSY), \cite{Schmidt},..., etc..

There are various experimental results on the fermion EDMs in the
literature and they read, $d_e =(1.8\pm 1.2\pm 1.0)\times 10^{-27}
e\, cm$ \cite{Commins}, $d_{\mu} =(3.7\pm 3.4)\times 10^{-19} e\,
cm$ \cite{Bailey} and $|d_{\tau}| <(3.1)\times 10^{-16} e\, cm$
\cite{Groom} respectively.

From the theoretical point of view, the extensive work has been
done on the EDMs of fermions in various models, in the literature.
The lepton electric dipole moments have been studied in the seesaw
model, in \cite {Bhaskar}. The electron EDM has been predicted at
the order of $10^{-32}\, e-cm$, in the 2HDM, including tree level
flavor changing neutral currents (FCNC) and complex Yukawa
couplings, in \cite {Iltmuegam}. The work \cite{IltanNonCom} is
devoted to the fermion EDM moments in the SM with the inclusion of
non-commutative geometry. In \cite{IltanExtrEDM}
(\cite{IltanSplitEDM}), the electric dipole moments of fermions in
the two Higgs doublet model with the inclusion of non-universal
extra dimensions (in the split fermion scenario) have been
analyzed. On the other hand, the quark EDMs have been estimated in
several models \cite{sahab1} and the EDMs of nuclei, deutron,
neutron and some atoms have been predicted extensively
\cite{Vladimir}.

In the present work, we study the lepton EDMs in the two Higgs
doublet model in which the flavor changing (FC) neutral current
vertices in the tree level are permitted and the CP violating
interactions are carried by complex Yukawa couplings. In the
calculations, we include the effects of a possible new dimension,
in the split fermion scenario, where the hierarchy of fermion
masses are coming from the overlap of the fermion Gaussian
profiles in the extra dimension. In addition to the strong
localization of the leptons, we consider that the new Higgs
scalars are also localized around the the origin, in the extra
dimension. These localizations are the result of the non vanishing
couplings between leptons (new Higgs scalars) and the so called
localizer field, which is odd under $Z_2$ reflection in the extra
dimension. The split fermion scenario has been studied in several
works in the literature \cite{Hamed}-\cite{Surujon}. In
\cite{Hamed}, the fermion mass hierarchy has been introduced by
assuming that the fermions were located at different points in the
extra dimensions, with exponentially small overlaps of their wave
functions. The phenomenologically reliable locations of left and
right handed components of fermions in the extra dimensions and
their roles on the mechanism of Yukawa hierarchies have been
studied in \cite{Mirabelli}.  \cite{Chang} was devoted to
prediction of the constraint on the split fermions in the extra
dimensions  by considering leptonic W decays and the lepton
violating processes. The CP violation in the quark sector in the
split fermion model was studied in \cite{Branco}  and in
\cite{Chang2}, the new configuration of split fermion positions in
a single extra dimension and the physics of kaon, neutron and B/D
mesons  have been analyzed to find stringent bounds on the size of
the compactification scale 1/R. The shapes and overlaps of the
fermion wave functions in the split fermion model have been
studied in \cite{Perez} and the rare processes in the split
fermion scenario have been analyzed in \cite{Hewett}.
\cite{IltanSplitEDM} (\cite{IltanLFVSplit},\cite{IltanZl1l2Split})
was related to the electric dipole moments of charged leptons (the
radiative lepton flavor violating  (LFV) decays, the LFV
$Z\rightarrow l_i\,l_j$ decays) in the split fermion scenario. The
work \cite{IltanLFVSplitFat} was devoted to the branching ratios
of the radiative LFV decays in the split fermion scenario, with
the assumption that the new Higgs doublet is restricted to the 4D
brane or to the thin bulk  in one and two extra dimensions, in the
framework of the two Higgs doublet model. Recently, the Higgs
localization in the split fermion models has been studied in
\cite{Surujon}.

The present work is devoted to the EDMs of charged leptons in the
2HDM with the assumption that they have Gaussian profiles in the
extra dimension, similar to the previous work
\cite{IltanSplitEDM}. However, we also consider that the new Higgs
scalars are localized in the extra dimension with the help of the
so called localizer field. The idea of the localization of the SM
Higgs, using the localizer field, has been studied in
\cite{Surujon}. Here, we assume that the new Higgs scalars are
localized in the extra dimension and the SM Higgs has a constant
profile. The localization of the new Higgs scalars depends
strongly on the strength of the coupling of the localizer to the
new Higgs scalar. Here we take this coupling small and study the
sensitivity of the lepton EDMs on it. Furthermore, we analyze the
compactification scale dependence of lepton EDMs. In the numerical
calculations, we observe that the lepton EDMs are sensitive to the
new Higgs localization in the extra dimension and we estimate that
the numerical values of electron (muon, tau) EDM is at the order
of the magnitude of $10^{-31}\, (10^{-24},\, 10^{-22})$ e-cm, even
for the large values of the compactification scale $1/R$ and for
the intermediate values of Yukawa couplings.

The paper is organized as follows: In Section 2, we present EDMs
of charged leptons in the split fermion scenario, with the
localization of new Higgs scalars in the extra dimension, in the
2HDM. Section 3 is devoted to discussion and our conclusions. In
Appendix section we present the derivation of the KK modes of new
Higgs fields and their masses in the case that they are coupled to
the localizer field.
\section{The effect of the localization of new Higgs scalars on the
electric dipole moments of charged leptons, in the split fermion
scenario, in the two Higgs doublet model. }
The existence of the fermion EDM is the signal for the CP
violation, since it is the result of the CP violating
fermion-fermion-photon effective interaction. In the framework of
the SM, the complex CKM matrix (possible lepton mixing matrix) is
the possible source of this violation for quarks (for charged
leptons). Since the theoretical estimated numerical values of the
fermion EDMs are considerably small, there is a need to go beyond
the SM to get additional contributions, in order to enhance their
numerical values. The multi Higgs doublet models are the among the
candidates for this enhancement, and, in the present work, we take
the 2HDM, which allows the FCNC at tree level and includes the
complex Yukawa couplings as a source of CP violation. On the other
hand, with the addition of extra dimensions, there exist new
contributions sensitive to the compactification scale $1/R$ where
$R$ is the compactification radius of the extra dimension. Here,
we take the effects of extra dimensions into account with the
assumption that the hierarchy of lepton masses are coming from the
lepton Gaussian profiles in the extra dimensions, so called the
split fermion scenario. This localization is obtained by the
nonzero coupling of the localizer scalar field $\phi_L$ to the
lepton fields. Furthermore, we assume that this field couples also
to the new Higgs doublet and localizes the new Higgs scalars
around the origin $y=0$, where $y$ is the coordinate of the extra
dimension (see Appendix).

The Yukawa Lagrangian which creates the lepton EDM in a single
extra dimension, respecting the split fermion scenario, reads:
\begin{eqnarray}
{\cal{L}}_{Y}=
\xi^{E}_{5 \,ij} \bar{\hat{l}}_{i L} \phi_{2} \hat{E}_{j R} + h.c.
\,\,\, , \label{lagrangian}
\end{eqnarray}
where $L$ and $R$ denote chiral projections $L(R)=1/2(1\mp
\gamma_5)$, $\phi_{2}$ is the new scalar doublet. Here $\hat{l}_{i
L}$ ($\hat{E}_{j R}$), with family indices $i,j$, are the zero
mode \footnote{In our calculations, we take only zero mode lepton
fields.} lepton doublets (singlets) with Gaussian profiles in the
extra dimension $y$ and they read
\begin{eqnarray}
\hat{l}_{i L}&=& N\,e^{-(y-y_{i L})^2/2 \sigma^2}\,l_{i L} ,
\nonumber
\\ \hat{E}_{j R}&=&N\, e^{-(y-y_{j R})^2/2 \sigma^2}\, E_{j R}\, ,
\label{gaussianprof}
\end{eqnarray}
with the normalization factor $N=\frac{1}{\pi^{1/4}\,
\sigma^{1/2}}$. $l_{i L}$ ($E_{j R}$) are the lepton doublets
(singlets) in four dimensions. Here the parameter $\sigma$ is the
Gaussian width of the leptons with the property $\sigma << R$, and
$y_{i (L,R)}$ are the fixed position of $i^{th}$ left (right)
handed lepton in the fifth dimension\footnote{The positions of
left handed and right handed leptons are obtained by taking the
observed masses into account \cite{Mirabelli}. The idea is that
the lepton mass hierarchy is due to the the relative positions of
the Gaussian peaks of the wave functions located in the extra
dimension \cite{Hamed, Mirabelli}. By assuming that the lepton
mass matrix is diagonal, one possible set of locations for the
lepton fields read (see \cite{Mirabelli} for details)
\begin{eqnarray}
P_{l_i}=\sqrt{2}\,\sigma\, \left(\begin{array}{c c c}
11.075\\1.0\\0.0
\end{array}\right)\,,\,\,\,\, P_{e_i}=\sqrt{2}\,\sigma\, \left(\begin{array}
{c c c} 5.9475\\4.9475\\-3.1498
\end{array}\right)
 \,\, . \label{location}
\end{eqnarray}}.

The Higgs doublets $\phi_{1}$ and $\phi_{2}$ are chosen as
\begin{eqnarray}
\phi_{1}=\frac{1}{\sqrt{2}}\left[\left(\begin{array}{c c}
0\\v+H^{0}\end{array}\right)\; + \left(\begin{array}{c c} \sqrt{2}
\chi^{+}\\ i \chi^{0}\end{array}\right) \right]\, ;
\phi_{2}=\frac{1}{\sqrt{2}}\left(\begin{array}{c c} \sqrt{2}
H^{+}\\ H_1+i H_2 \end{array}\right) \,\, . \label{choice}
\end{eqnarray}
so that their vacuum expectation values read:
\begin{eqnarray}
<\phi_{1}>=\frac{1}{\sqrt{2}}\left(\begin{array}{c c}
0\\v\end{array}\right) \,  \, ; <\phi_{2}>=0 \,\, .
\label{choice2}
\end{eqnarray}
This leads to the possibility of collecting SM (new) particles in
the first (second) doublet. In this case $H_1$ and $H_2$ are the
mass eigenstates $h^0$ and $A^0$ respectively since no mixing
occurs between two CP-even neutral bosons $H^0$ and $h^0$ at tree
level.\footnote{Here we consider the Higgs potential term
$V(\phi_1, \phi_2,\phi_L)$ as,
\begin{eqnarray}
V(\phi_1, \phi_2,\phi_L)=c_1 (\phi_1^+ \phi_1-v^2/2)^2+ c_2
(\phi_2^+ \phi_2)^2+
\frac{1}{2}\,g\,\,\phi_2^{\dagger}\,\phi_2\,\phi_L^2 \, ,
\label{Pot11}
\end{eqnarray}
with constants $c_i, \, i=1, 2$, where $\phi_L$ is the localizer
singlet field (see Appendix section for details). In this case,
the first Higgs doublet does not couple to the second one and to
the localizer singlet field. With this assumption, the two Higgs
doublets do not mix and the SM (new) particles are placed in the
first (second) doublet. If the first and the second doublets
couple and the second one acquires non zero vacuum expectation
value, the mixing appears between two doublets and it needs
further analysis.}

Now we take that the SM Higgs has a constant profile in the extra
dimension and the new Higgs field $\phi_{2}$, which has the main
role in the existence of the charged lepton EDM, couples to the
localizer $\phi_L$ with a small coupling $g$ (see the Appendix
section for details). The coupling of the new Higgs doublet to the
localizer results in the localization of the Higgs scalars, with
the corresponding KK modes, around the origin, and brings modified
Yukawa interactions in four dimensions. The neutral CP even and
odd scalar fields S ($S=h^0, \,A^0$), are expanded into their KK
modes as  $S(x,y)=\sum_n h_n(y)\,S^{(n)}(x)$ (see the Appendix
section) and to obtain the lepton-lepton-Higgs interaction
coupling in four dimensions we need to integrate the combination
$\bar{\hat{f}}_{iL\,(R)}\,S^{(n)}(x)\,h_n(y)\, \hat{f}_{j
R\,(L)}$, appearing in the part of the Lagrangian (eq.
(\ref{lagrangian})), over the fifth dimension. Using the KK basis
obtained (see eq. (\ref{f0}) and (\ref{fn})), we get
\begin{eqnarray}
\int_{-\pi R}^{\pi R}\, dy\,\,
\bar{\hat{f}}_{iL\,(R)}\,S^{(n)}(x)\,h_n (y)\,
\hat{f}_{jR\,(L)}=V^n_{LR\,(RL)\,ij} \, \bar{f}_{i L\, (R)}
\,S^{(n)} (x)\,\,f_{j R\,(L)}\,\, , \label{intVij1}
\end{eqnarray}
where the factor $V^n_{LR\,(RL)\,ij}$ reads
\begin{eqnarray}
V^n_{LR\,(RL)\,ij}=V^0_{LR\,(RL)\,ij} \, c^{(')}_n\,(i,j) \, ,
\label{Vij1even}
\end{eqnarray}
and the function $V^0_{LR\,(RL)\,ij}$ is
\begin{eqnarray}
V^0_{LR\,(RL)\,ij}=\frac{g^{1/8}\,e^{-(y_{i L\, (R)}-y_{j R\,
(L)})^2/4
\sigma^2}}{\sqrt{\sigma}\,\pi^{1/4}\,\sqrt{Erf[\pi\,g^{1/4}\,
\frac{R}{\sigma}]}}\, .
\end{eqnarray}
Here, the fields $f_{iL}$, $f_{j R}$ are the four dimensional
lepton fields and  the function $Erf[z]$ is the error function,
which is defined as
\begin{eqnarray}
Erf[z]=\frac{2}{\sqrt{\pi}}\,\int_{0}^{z}\,e^{-t^2}\,dt \,\, .
\label{erffunc}
\end{eqnarray}
The functions $c^{(')}_n (i,j)$ in eq. (\ref{Vij1even}) are
calculated in the case that the coupling $g$ is non zero and
small:
\begin{eqnarray}
c^{(')}_ n(i,j)&=&\!\!\!\!\frac{ (-1)^n\,2^{3
n+\frac{1}{2}}\,\sqrt{\pi}\, (1-\frac{g}{4})^n\,
(2+\sqrt{g})^{-\frac{4n+1}{2}}} {\sqrt{(2\,n)!}\,\,
\Gamma[\frac{1}{2}-n]}\,e^{f_{LR\,(RL)\,ij}} \nonumber \\
&\times&
F_{\!\!\!\!\!\!\!\!1\,\,\,\,\,1}[-n,\frac{1}{2},\frac{\sqrt{g}\,(y_{i
L\, (R)}+y_{j R\, (L)})^2} {(4-g)\,\sigma^2}] \label{cevenodd} \,
,
\end{eqnarray}
where
\begin{eqnarray}
f_{LR\,(RL)\,ij}=-\frac{\sqrt{g}\,(y_{i L\, (R)}+y_{j R\,
(L)}))^2} {4\,(2+\sqrt{g})\,\sigma^2} \label{fLR}\, .
\end{eqnarray}
The function \, $F_{\!\!\!\!\!\!\!\!1\,\,\,\,\,1} [a;b;z]$
appearing in eq. (\ref{cevenodd}) is the hypergeometric function
\begin{eqnarray}
F_{\!\!\!\!\!\!\!\!1\,\,\,\,\,1} [a;b;z]=\sum_{k=0}^{\infty}\,
(a)_k\, (b)_k\,z^k/k! \label{HyperGeo}\, ,
\end{eqnarray}
where $(d)_k= \frac{\Gamma[d+k]}{\Gamma[d]}$.
%
%
Finally, we can define the Yukawa couplings in four dimensions as
\begin{eqnarray}
\xi^{E}_{ij}\,\Big((\xi^{E}_{ij})^\dagger\Big)= V^0_{LR\,(RL)\,ij}
\, \xi^{E}_{5\, ij}\,\Big((\xi^{E}_{5\, ij})^\dagger\Big)\, ,
\label{coupl4}
\end{eqnarray}
where $\xi^{E}_{5\, ij}$ are  the Yukawa couplings in five
dimensions (see eq. (\ref{lagrangian}))
\footnote{In the following we use the dimensionful coupling
$\bar{\xi}^{E}_{N}$ in four dimensions, with the definition
$\xi^{E}_{N,ij}=\sqrt{\frac{4\, G_F}{\sqrt{2}}}\,
\bar{\xi}^{E}_{N,ij}$ where N denotes the word "neutral".}.

\begin{figure}[htb]
\vskip 0.5truein \centering \epsfxsize=2.8in
\leavevmode\epsffile{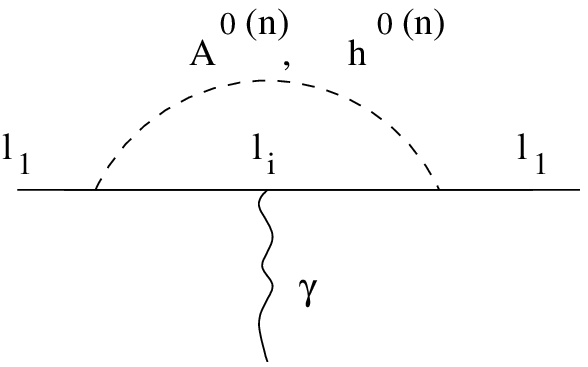} \vskip 0.5truein \caption[]{One loop
diagrams contribute to EDM of charged leptons due to neutral Higgs
bosons $h^0$, $A^0$ in the 2HDM, including KK modes in a single
extra dimension. Wavy lines represent the electromagnetic field
and dashed lines the Higgs field where $l_{1 \,(i)}=e, \mu, \tau$}
\label{fig1}
\end{figure}

The effective EDM interaction for a charged lepton $f$ is given by
\begin{eqnarray}
{\cal L}_{EDM}=i d_f \,\bar{f}\,\gamma_5 \,\sigma^{\mu\nu}\,f\,
F_{\mu\nu} \,\, , \label{EDM1}
\end{eqnarray}
where $F_{\mu\nu}$ is the electromagnetic field tensor, '$d_{f}$'
is EDM of the charged lepton and it is a real number by
hermiticity. With the assumption that the CP violating EDM
interaction comes from the complexity of the Yukawa couplings due
to the new Higgs scalars \footnote{We do not consider the possible
effects due to the CKM type lepton mixing matrix and take only
zero mode lepton fields.}, the $f$-lepton EDM '$d_f$'
$(f=e,\,\mu,\,\tau)$  can be calculated as a sum of contributions
coming from neutral Higgs bosons $h_0$ and $A_0$ (see Fig.
\ref{fig1}),
\begin{eqnarray}
d_f&=& -\frac{i\, G_F}{\sqrt{2}} \frac{e}{32\pi^2}\,
\frac{Q_{\tau}}{m_{\tau}}\, ((\bar{\xi}^{E\,*}_{N,l\tau})^2-
(\bar{\xi}^{E}_{N,\tau l})^2)\, \Bigg ( c_0 (f,\tau)\,c'_0
\,(f,\tau)\,(F_1 (y_{h_0})-F_1 (y_{A_0}))\nonumber \\  &+&
\!\!\!\! \sum_{n=1}^{\infty}\,  c_n (f,\tau)\,c'_n (f,\tau)\, (F_1
(y^{n}_{h_0})-F_1 (y^{n}_{A_0})) \Bigg) , \nonumber
\\  \label{emuEDM}
\end{eqnarray}
for $f=e,\mu$ and
\begin{eqnarray}
d_{\tau}&=& -\frac{i\,G_F}{\sqrt{2}} \frac{e}{32\pi^2}\, \Big{\{}
\frac{Q_{\tau}}{m_{\tau}}\, ((\bar{\xi}^{E\,*}_{N,\tau\tau})^2-
(\bar{\xi}^{E}_{N,\tau \tau})^2)\, \Bigg( c_0^2 (\tau,\tau)\,(F_2
(r_{h_0})-F_2(r_{A_0}))\nonumber \\ &+& \sum_{n=1}^{\infty}\,c_n^2
(\tau,\tau)\,(F_2 (r^{n}_{h_0})-F_2(r^{n}_{A_0})) \Bigg) \nonumber
\\&-&  Q_{\mu}\, \frac{m_{\mu}}{m^2_{\tau}}\,
((\bar{\xi}^{E\,*}_{N,\mu\tau})^2-(\bar{\xi}^{E}_{N,\tau
\mu})^2)\, \Bigg( c_0 (\mu,\tau)\,c'_0 (\mu,\tau)\,(r_{h_0}\,ln\,
(z_{h_0})-r_{A_0}\,ln\, (z_{A_0}))\nonumber \\ &+&
\sum_{n=1}^{\infty}\,c_n (\mu,\tau)\,c'_n (\mu,\tau)\,
(r^{n}_{h_0}\,ln\, (z^{n}_{h_0})-r^{n}_{A_0}\,ln\, (z^{n}_{A_0}))
\Bigg) \Big{\}} \,\, , \label{tauEDM}
\end{eqnarray}
for $f=e,\mu, \tau$. Here the functions $F_1 (w)$, $F_2 (w)$ read
\begin{eqnarray}
F_1 (w)&=&\frac{w\,(3-4\,w+w^2+2\,ln\,w)}{(-1+w)^3}\nonumber \,\, , \\
F_2 (w)&=& w\, ln\,w + \frac{2\,(-2+w)\, w\,ln\,
\frac{1}{2}(\sqrt{w}-\sqrt{w-4})}{\sqrt{w\,(w-4)}} \, ,
\label{functions1}
\end{eqnarray}
with $y_{S}=\frac{m^2_{\tau}}{m^2_{S}+2\,\beta}$,
$y^{n}_{S}=\frac{m^2_{\tau}}{m^2_{S}+2\,(4 n+1)\,\beta}$,
$r_{S}=\frac{1}{y_{S}}$, $r^{n}_S=1/y^{n}_{S}$,
$z_{S}=\frac{m^2_{\mu}}{m^2_{S}+2\,\beta}$,
$z^{n}_{S}=\frac{m^2_{\mu}}{m^2_{S}+2\,(4 n+1)\,\beta}$,
$Q_{\tau}$, $Q_{\mu}$ are charges of $\tau$ and $\mu$ leptons
respectively. Notice that the functions $c^{(')}_
n(i,j)$\footnote{If the second Higgs doublet is localized around
$y=y_0$, the zero mode solution reads $h_0(y)=N'\, e^{-\beta\,
(y-y_0)^2}$ (see eq. (\ref{f0})), and, if $y_0$ is away from the
origin, $y_0\rightarrow \pi\, R$, the coefficients $c^{(')}_
n(i,j)$ tends to zero. This is the case where the new Higgs
contribution vanishes.} are defined in eq. (\ref{cevenodd}). In
eq. (\ref{emuEDM}) we take into account only the internal
$\tau$-lepton contribution respecting our assumption that the
Yukawa couplings $\bar{\xi}^{E}_{N, ij},\, i,j=e,\mu$, are small
compared to $\bar{\xi}^{E}_{N,\tau\, i}\, i=e,\mu,\tau$ due to the
possible proportionality of the Yukawa couplings to the masses of
leptons in the vertices. In eq. (\ref{tauEDM}), we present also
the internal $\mu$-lepton contribution, which can be neglected
numerically. Notice that, we make our calculations in arbitrary
photon four momentum square $q^2$ and take $q^2=0$ at the end.

Finally, in our calculations, we choose the Yukawa couplings
complex and we used the parametrization
\begin{eqnarray}
\bar{\xi}^{E}_{N,\tau f}=|\bar{\xi}^{E}_{N,\tau f}|\, e^{i\,
\theta_{f}} \,. \label{xi2}
\end{eqnarray}
Therefore, the Yukawa factors in eqs. (\ref{emuEDM}) and
(\ref{tauEDM}) can be written as
\begin{eqnarray}
((\bar{\xi}^{E\,*}_{N,f\tau})^2-(\bar{\xi}^{E}_{N,\tau
f})^2)=-2\,i \,sin\,2\theta_{f}\, |\bar{\xi}^{E}_{N,\tau f}|^2
\end{eqnarray}
where $f=e,\mu,\tau$. Here $\theta_{f}$ is the CP violating
parameter which is the source of the lepton EDM.

\section{Discussion}
The fermion EDM is the result of the existence of the CP violation
and arises from the fermion-fermion-photon effective interaction.
In the present work, we study the charged lepton EDMs in the 2HDM
with FCNC at tree level and we assume that the CP phase is coming
from the complex Yukawa couplings, driving the lepton-lepton-new
Higgs scalar vertices. Furthermore, we take into account the
effects of a single extra dimension in the so called split fermion
scenario where the hierarchy of lepton masses are coming from the
lepton Gaussian profiles in the extra dimension. In this scenario,
the penetration of leptons into the bulk is ensured by the
non-zero coupling of leptons with a background scalar field,
namely the localizer field, $\phi_L$, which has a vacuum
expectation value, depending on the coordinate of the extra
dimension and centered at the origin. In addition to this, we
consider that the new Higgs doublet also couples to the localizer
weakly and the corresponding Higgs scalars localize around the
origin $y=0$. On the other hand, the SM Higgs $H^0$ is taken as
uniform in the extra dimension and its vacuum expectation value is
responsible for the generation of the masses. Notice that the mass
term, which is modulated by the mutual overlap of lepton
wavefunctions, is obtained by integrating the operator
$H^0\,\bar{\hat{f}}\,\hat{f}$ over extra dimensions, where $f$
denotes lepton. This is the idea of fixing the positions of left
(right) handed leptons in the extra dimensions (see
\cite{Mirabelli} for details). Since the lepton-lepton-new scalar
vertices creates the CP violating EDM interaction, besides zero
modes, the KK modes of leptons and Higgs scalars have additional
contributions. Here, we include the effects of Higgs KK modes,
however, we ignored the lepton KK mode contributions to the EDMs
of leptons, by expecting that, for the large values of the
compactification scale, their effects on the physical parameters
are suppressed.

In the numerical calculations, the free parameters coming the
model used should be fixed by using the present experimental
results. The first set of free parameters is the four dimensional
leptonic complex couplings $\bar{\xi}^E_{N,ij}, i,j=e, \mu, \tau$.
We consider the Yukawa couplings $\bar{\xi}^{E}_{N,ij},\,
i,j=e,\mu $, as smaller compared to $\bar{\xi}^{E}_{N,\tau\, i}\,
i=e,\mu,\tau$ and we assume that $\bar{\xi}^{E}_{N,ij}$ is
symmetric with respect to the indices $i$ and $j$. In the case
that no extra dimension exists, the experimental uncertainty,
$10^{-9}$, in the measurement of the muon anomalous magnetic
moment \cite{BNL} can be used to estimate the upper limit of
$\bar{\xi}^{E}_{N,\tau \mu}$, since the new physics effects can
not exceed this uncertainty and this limit was predicted as $30\,
GeV$ (see \cite{Iltananomuon} and references therein) . Using this
upper limit and the experimental upper bound of BR of
$\mu\rightarrow e \gamma$ decay, BR $\leq 1.2\times 10^{-11}$, the
coupling $\bar{\xi}^{E}_{N,\tau e}$ is restricted in the range,
$10^{-3}-10^{-2}\,GeV$ (see \cite{Iltmuegam}). In our
calculations, we choose the numerical values of the couplings
$\bar{\xi}^{E}_{N,\tau \mu}$ ($\bar{\xi}^{E}_{N,\tau e}$) not
larger than $30\, GeV$ (around $10^{-3}\, GeV$). For the coupling
$\bar{\xi}^{E}_{N,\tau \tau}$, we use the numerical values which
are greater than $\bar{\xi}^{E}_{N,\tau \mu}$, since we have no
explicit restriction region. For the CP violating parameter which
drives the EDM interaction we choose the intermediate value,
$sin\,\theta_{e\, (\mu,\, \tau)}=0.5$

The second set of free parameters is coming from the split fermion
scenario where the hierarchy of lepton masses are due to the
lepton Gaussian profiles in the extra dimension. Here, we have a
free parameter  $\rho=\sigma/R$, where $\sigma$ is the Gaussian
width of the fermions and $R$ is the compactification radius.
There are various predictions on the compactification radius in
the literature. The lower bound for inverse of the
compactification radius was estimated as $\sim 300\, GeV$
\cite{Appelquist} in the case of universal extra dimension
scenario. The direct limits from searching for KK gauge bosons
imply $1/R> 800\,\, GeV$, the precision electro weak bounds on
higher dimensional operators generated by KK exchange place a far
more stringent limit $1/R> 3.0\,\, TeV$ \cite{Rizzo} and, from
$B\rightarrow \phi \, K_S$, the lower bounds for the scale $1/R$
have been obtained as $1/R > 1.0 \,\, TeV$, from $B\rightarrow
\psi \, K_S$ one got $1/R
> 500\,\, GeV$, and from the upper limit of the $BR$, $BR \, (B_s
\rightarrow \mu^+ \mu^-)< 2.6\,\times 10^{-6}$, the estimated
limit was $1/R > 800\,\, GeV$ \cite{Hewett}. On the other hand,
there exist different phenomenologically reliable set of locations
of lepton fields in the extra dimension, even if they are
predicted by using the lepton masses. Notice that we used the set
\footnote{The numerical results of physical parameters depends on
choice of the set of locations of the lepton fields.} predicted by
the work \cite{Mirabelli}. Finally, the coupling $g$, which exists
since we consider that the new Higgs doublet also couples to the
localizer, is another free parameter of the model used. Here, we
assume that this coupling is in a region such that the new Higgs
scalars are localized around the origin as much as possible.
Notice that the zero mode Higgs boson $h^0$ and $A^0$ masses
$m_{0,\,h^0\,(A^0)}^2=m_{h^0\,(A^0)}^2+2\,\beta$ (with $\beta=
\sqrt{g}\,\mu^2$) depend on the coupling $g$ and the additional
term comes from the localization effect, which should not be large
\footnote{Here we choose the masses coming from the Higgs mass
term $m_S$ in the lagrangian (see eq. (\ref{Pot1})) as
$m_{h^0}\sim 0\, (GeV)$ and $m_{A^0}\sim 180\, (GeV)$ and choose
the coupling $g$ in the range such that the zero mode masses
$m_{0,\,h^0\,(A^0)}$ are around the numerical values as
$100\,(200)\,GeV$.}.

Now, we start to estimate the charged lepton EDMs and study the
coupling $g$ and the parameter $\rho$ dependence of these physical
quantities.

\begin{figure}[htb]
\vskip -3.0truein \centering \epsfxsize=6.8in
\leavevmode\epsffile{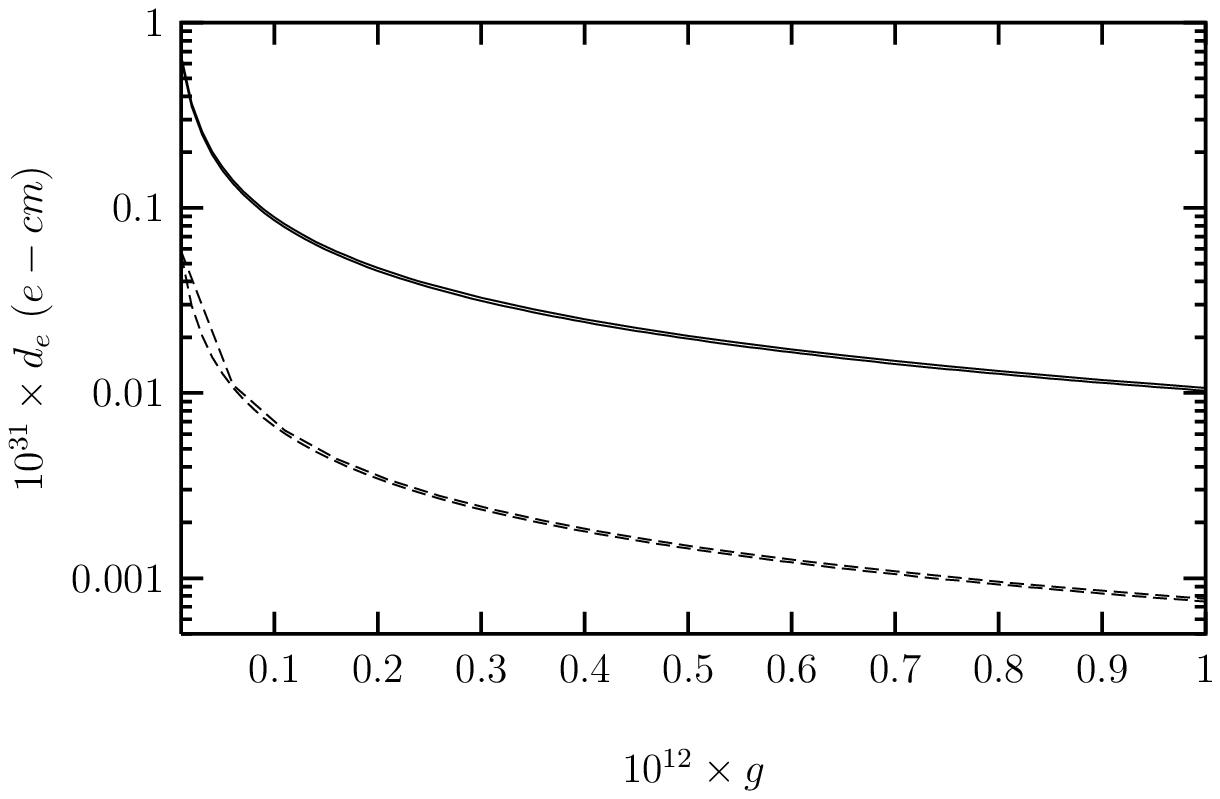} \vskip -3.0truein \caption[]{
$d_{e}$ with respect to the coupling $g$ for $1/R=1000\,GeV$,
$\bar{\xi}^{E}_{N,\tau e} =0.001\, GeV$. Here the lower-upper
solid (dashed) line represents the EDM, due to the zero mode
Higgs-KK mode Higgs included contribution for the parameter
$\rho=0.001$ ($\rho=0.0005$).} \label{EDMeg}
\end{figure}

In  Fig. \ref{EDMeg}, we plot the electron EDM $d_{e}$ with
respect to the coupling $g$ for $1/R=1000\,GeV$ and
$\bar{\xi}^{E}_{N,\tau e} =0.001\, GeV$. Here the lower-upper
solid (dashed) line represents the EDM, due to the zero mode
Higgs-KK mode Higgs included contribution for the parameter
$\rho=0.001$ ($\rho=0.0005$). The sensitivity of EDM to the
coupling $g$ is strong. The addition of KK mode contributions
enhance the EDM weakly, especially for the parameter $\rho=0.001$.
The electron EDM is at most at the order of the magnitude of
$10^{-31}-10^{-32}\, (e-cm)$, for $g\sim 10^{-14}$ and
$\rho=0.001$ ($\rho=0.0005$).  The enhancement in EDM for the
small values of the coupling $g$ (and also for the large values of
the width $\sigma$) is due the fact that the masses of zero (and
also KK modes) (see eq. (\ref {KKModemass})) are smaller compared
to the ones for the large values of the coupling $g$ (and also for
the small values of the width $\sigma$). In the case that all the
2HDM particles are confined into the 4D brane and the extra
dimension effects are switched off, the electron EDM $d_{e}$ is at
the order of $10^{-30}\, (e-cm)$, which is the numerical value to
be reached with smaller coupling $g$ and the weak localization of
the new Higgs doublet, in the present scenario.

\begin{figure}[htb]
\vskip -3.0truein \centering \epsfxsize=6.8in
\leavevmode\epsffile{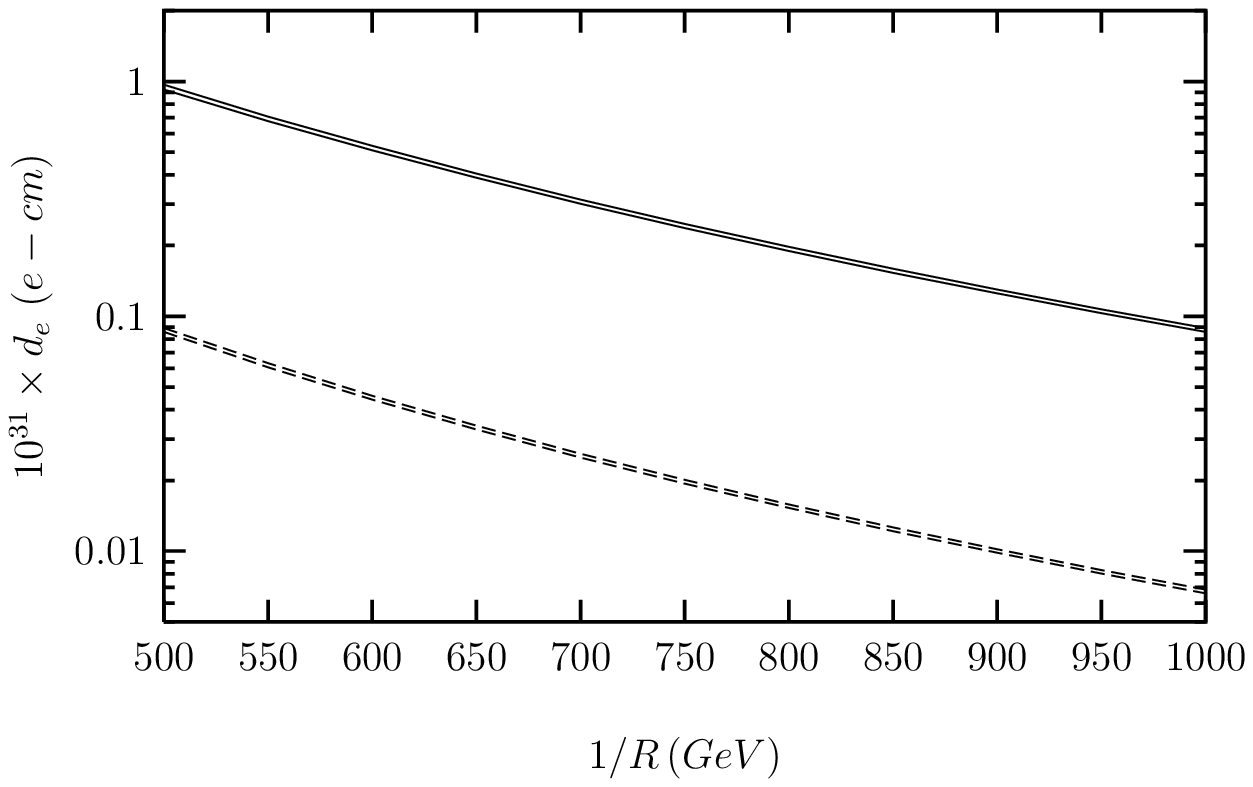} \vskip -3.0truein \caption[]{
$d_{e}$ with respect to $1/R$, for $g=10^{-13}$ and
$\bar{\xi}^{E}_{N,\tau e} =0.001\, GeV$. Here the lower-upper
solid (dashed) line represents the $d_{e}$, due to the zero mode
Higgs-KK mode Higgs included contribution for the parameter
$\rho=0.001$ ($\rho=0.0005$).} \label{EDMeR}
\end{figure}

Fig. \ref{EDMeR} is devoted to the compactification scale $1/R$
dependence of $d_{e}$ , for $g=10^{-13}$ and
$\bar{\xi}^{E}_{N,\tau e} =0.001\, GeV$. Here the lower-upper
solid (dashed) line represents the $d_{e}$, due to the zero mode
Higgs-KK mode Higgs included contribution for the parameter
$\rho=0.001$ ($\rho=0.0005$). The sensitivity of $d_{e}$ to $1/R$
is strong and it enhances with the decreasing values of $1/R$, as
expected. The effects of the addition the KK modes on $d_{e}$ are
weak.

\begin{figure}[htb]
\vskip -3.0truein \centering \epsfxsize=6.8in
\leavevmode\epsffile{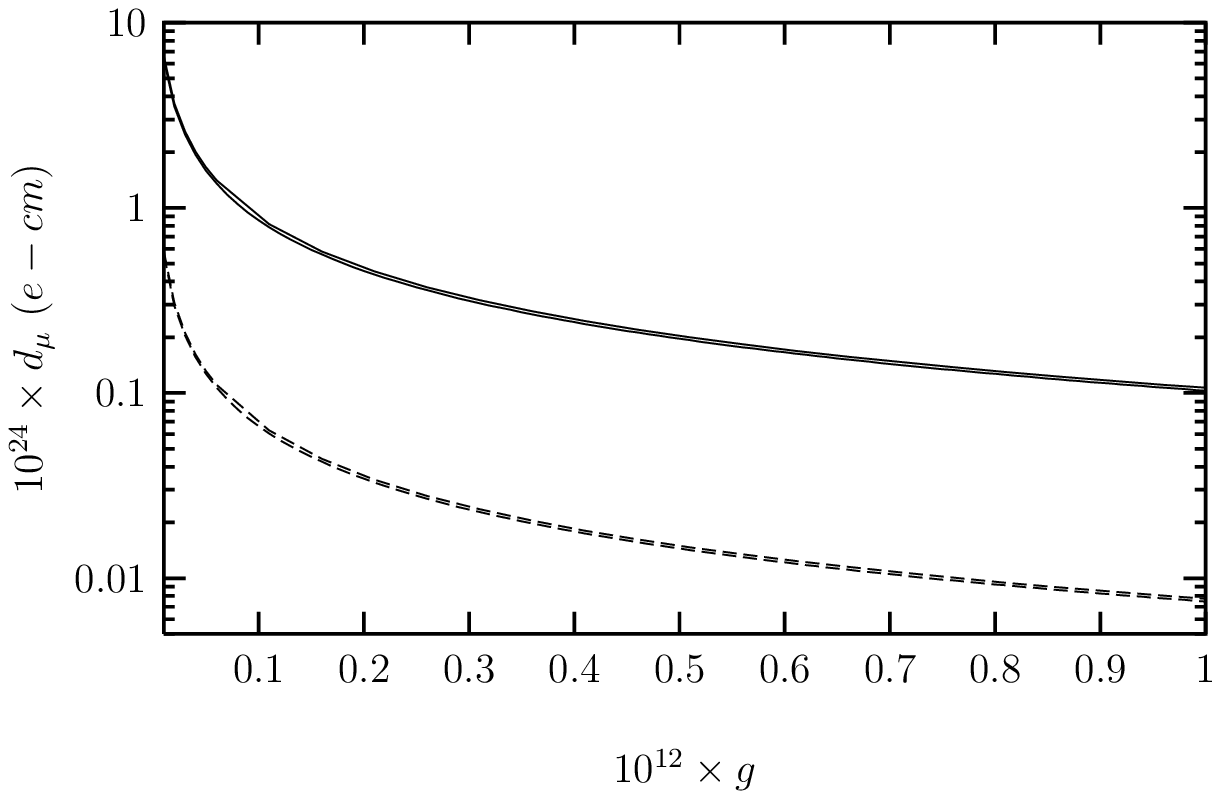} \vskip -3.0truein \caption[]{The
same as Fig. \ref{EDMeg} but for $d_{\mu}$ and
$\bar{\xi}^{E}_{N,\tau \mu} =10\, GeV$.} \label{EDMmug}
\end{figure}

\begin{figure}[htb]
\vskip -3.0truein \centering \epsfxsize=6.8in
\leavevmode\epsffile{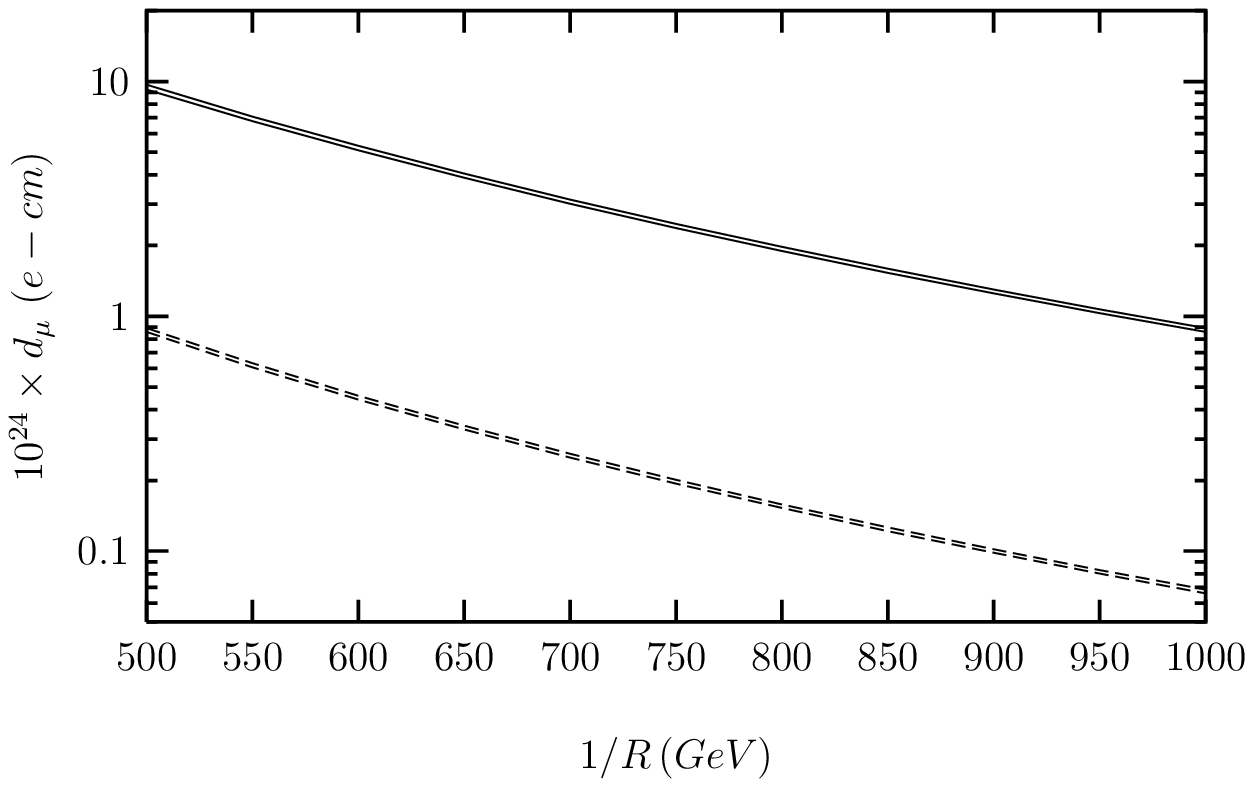} \vskip -3.0truein \caption[]{The
same as Fig. \ref{EDMeR} but for $d_{\mu}$ and
$\bar{\xi}^{E}_{N,\tau \mu} =10\, GeV$.} \label{EDMmuR}
\end{figure}

In  Fig. \ref{EDMmug}, we present $d_{\mu}$ with respect to the
coupling $g$ for $1/R=1000\,GeV$, $\bar{\xi}^{E}_{N,\tau \mu}
=10\, GeV$. Here the lower-upper solid (dashed) line represents
the EDM, due to the zero mode Higgs-KK mode Higgs included
contribution for the parameter $\rho=0.001$ ($\rho=0.0005$).
Similar to the electron EDM, the sensitivity of $d_{\mu}$ to the
coupling $g$ is strong and the addition of KK modes of new Higgs
scalars ensures negligible enhancement in the numerical value of
$d_{\mu}$. The muon EDM is at the order of $10^{-23}-10^{-24}\,
(e-cm)$, for $g\sim 10^{-14}$ and $\rho=0.001$ ($\rho=0.0005$).
The $1/R$ dependence of $d_{\mu}$ is presented Fig. \ref{EDMmuR},
for $g=10^{-13}$ and $\bar{\xi}^{E}_{N,\tau e} =0.001\, GeV$. Here
the lower-upper solid (dashed) line represents the $d_{\mu}$, due
to the zero mode Higgs-KK mode Higgs included contribution for the
parameter $\rho=0.001$ ($\rho=0.0005$). The sensitivity of
$d_{\mu}$ to $1/R$ is strong similar to the $d_{e}$ case. If the
extra dimension effects are switched off, the muon EDM $d_{\mu}$
is at the order of $10^{-22}\, (e-cm)$ and this numerical value
can be obtained with the weak localization of the new Higgs
doublet, in the present scenario.

\begin{figure}[htb]
\vskip -3.0truein \centering \epsfxsize=6.8in
\leavevmode\epsffile{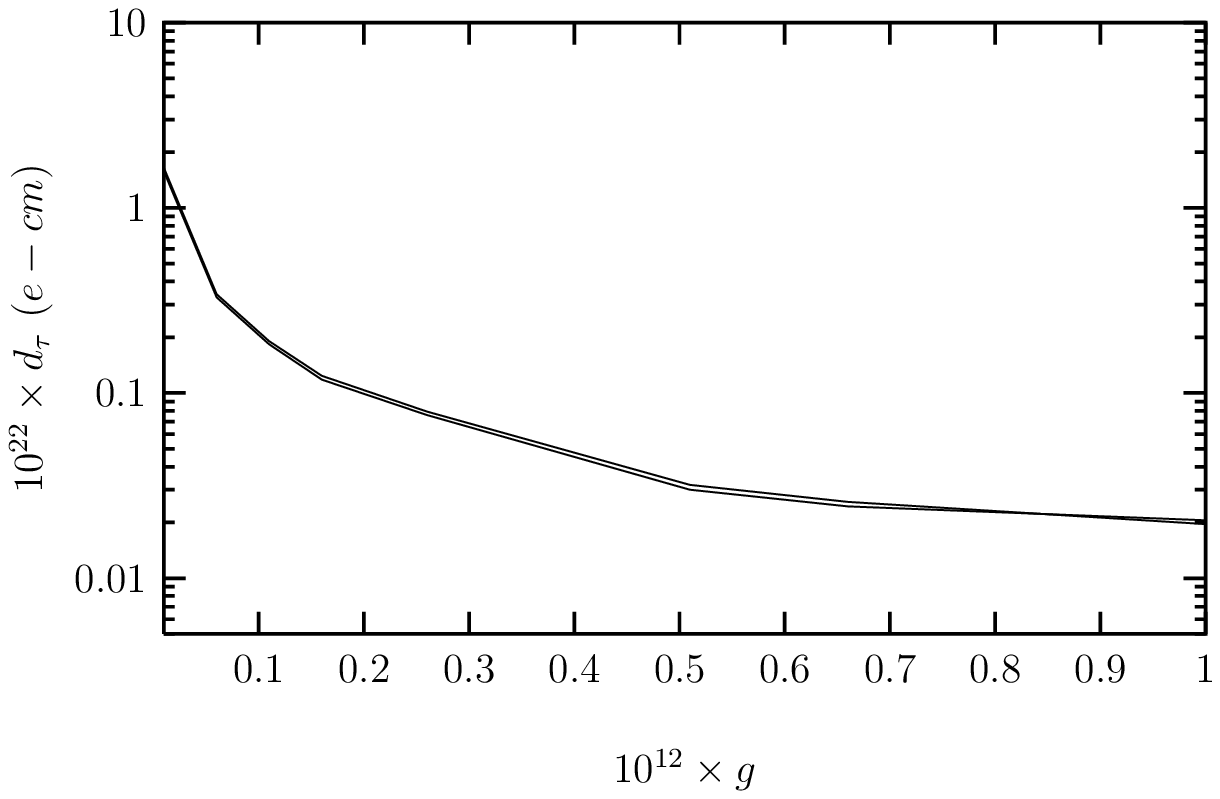} \vskip -3.0truein \caption[]{The
same as Fig. \ref{EDMeg} but for $d_{\tau}$ and $\rho=0.001$,
$\bar{\xi}^{E}_{N,\tau \mu} =10\, GeV$ and $\bar{\xi}^{E}_{N,\tau
\tau} =50\, GeV$.} \label{EDMtaug}
\end{figure}

\begin{figure}[htb]
\vskip -3.0truein \centering \epsfxsize=6.8in
\leavevmode\epsffile{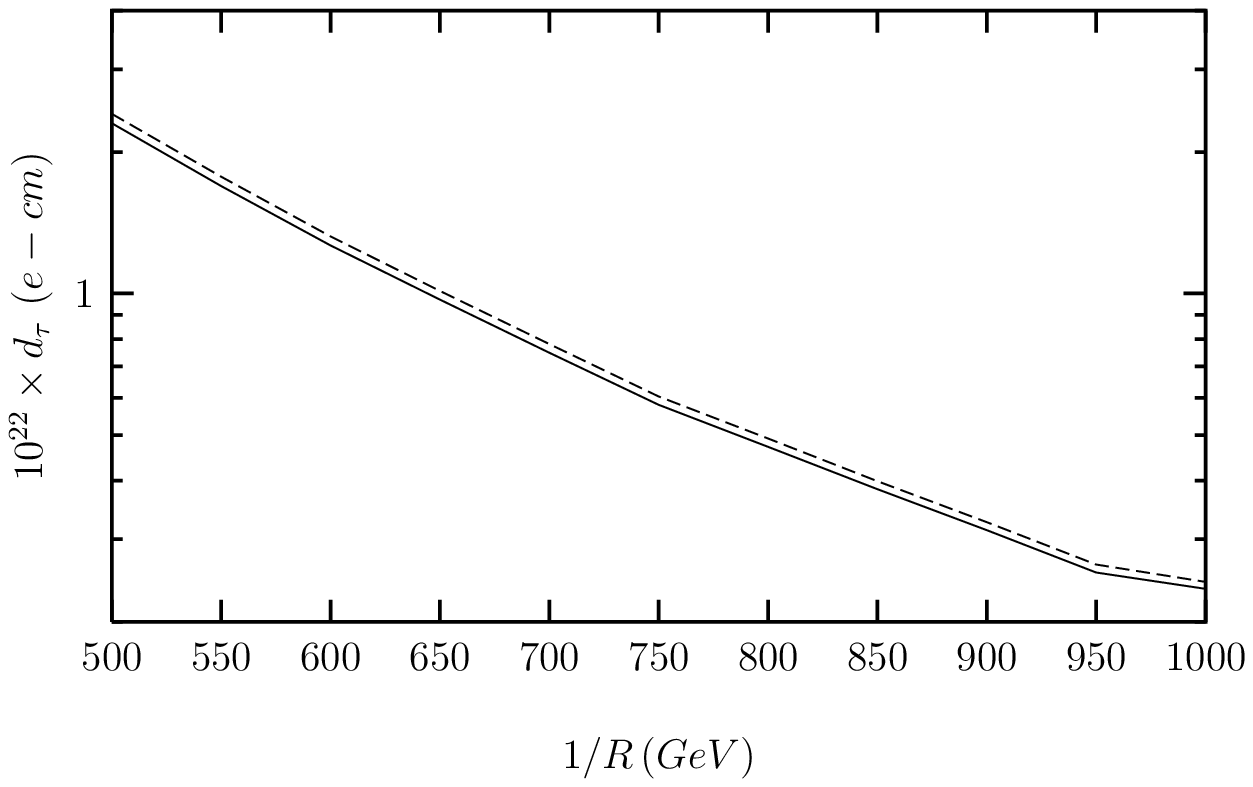} \vskip -3.0truein \caption[]{The
same as Fig. \ref{EDMeR} but for $d_{\tau}$ and $\rho=0.001$,
$\bar{\xi}^{E}_{N,\tau \mu} =10\, GeV$ and $\bar{\xi}^{E}_{N,\tau
\tau} =50\, GeV$.} \label{EDMtauR}
\end{figure}

Finally, we study the tau EDM $d_{\tau}$ in  Fig. \ref{EDMtaug}
and Fig. \ref{EDMtauR}. Fig. \ref{EDMtaug},  represents $d_{\tau}$
with respect to the coupling $g$ for $1/R=1000\,GeV$,
$\bar{\xi}^{E}_{N,\tau \mu} =10\, GeV$ and $\bar{\xi}^{E}_{N,\tau
\tau} =50\, GeV$. Here the lower-upper solid line represents the
EDM, due to the zero mode Higgs-KK mode Higgs included
contribution for the parameter $\rho=0.001$. We observe that
$d_{\tau}$ enhances up to the values at the order of $2\times
10^{-22} (e-cm)$ for $g\sim 10^{-14}$ and $\rho=0.001$. The
enhancement due to the KK modes of new Higgs scalars is
negligible. Fig. \ref{EDMtauR} is devoted to the $1/R$ dependence
of $d_{\tau}$ for $g=10^{-13}$ and $\bar{\xi}^{E}_{N,\tau e}
=0.001\, GeV$. Here the solid (dashed) line represents the
$d_{\tau}$, due to the zero mode Higgs (KK mode Higgs included)
contribution for the parameter $\rho=0.001$. The sensitivity of
$d_{\tau}$ to $1/R$ is strong and almost one order enhancement
with the decreasing values of $1/R$ from $1000\, (GeV)$ to $500\,
(GeV)$. The tau EDM $d_{\tau}$ is at the order of $10^{-21}\,
(e-cm)$ in the case that the extra dimension effects are switched
off. Such a numerical value is obtained for the weak localization
of the new Higgs doublet, similar to the electron and muon EDM
case.
\\

Now we would like to summarize our results:

\begin{itemize}
\item  The lepton EDMs are weakly sensitive the KK mode
contributions in the given range of the coupling $g$, however this
sensitivity increases for the small values of this coupling.
Furthermore, the numerical values of EDMs increase with the
decreasing values of the coupling $g$ and increasing values of
parameter $\rho$, which measures the Gaussian widths of leptons in
the extra dimension.
\item The sensitivity of lepton EDMs to $1/R$ increases with the
decreasing values of the parameter $\rho$.
\item  We obtain the numerical values at the order of $\sim
10^{-31} \,(10^{-24}, 10^{-21})\, (e-cm)$ for $g\sim 10^{-14}$ and
$\rho=0.001$, for the EDMs $d_{e}$ ($d_{\mu}, d_{\tau})$,
respectively. To be able to reach the results near the current
experimental limits, the following domains for the numerical
values of the various free parameters should be considered:
$\bar{\xi}^{E}_{N,\tau e}> 0.001\, GeV$, $\bar{\xi}^{E}_{N,\tau
\mu} > 10\, GeV$, $\bar{\xi}^{E}_{N,\tau \tau} > 50\, GeV$, $1/R<
500\, GeV$, $\rho > 0.001$ and $g\sim 10^{-14}$. It is obvious
that these theoretical values of the EDMs of charged leptons are
still far from the experimental limits. Furthermore, the bound for
the compactification scale almost coincides with the following
bounds existing in the literature: $1/R\sim 300\, GeV$
\cite{Appelquist} in the case of universal extra dimension
scenario and $1/R > 500\,\, GeV$ from $B\rightarrow \psi \, K_S$
\cite{Hewett}. Hopefully, the forthcoming more sensitive
experimental measurements will ensure smaller numerical values for
the lepton EDMs and, therefore,  the possible range of the free
parameter set will be more accurate.
\end{itemize}

As a final word, the future experimental measurements on leptons
EDMs would ensure valuable information about the possibility of
the existence of localizations of leptons and Higgs bosons in the
extra dimensions and make it more clear how the nature behaves
beyond the SM.
%
\section{Appendix}
\subsection{The KK modes of Higgs fields and their masses}
Here, we calculate the KK modes of Higgs fields and their masses
in the case that the localizer field $\phi_L$ couples to the new
Higgs fields weakly and localize these new fields around $y=0$,
where $y$ is the coordinate of the extra dimension. Our the
starting point is the Lagrangian of the new scalar field, which is
coupled to the localizer field:
\begin{eqnarray}
\emph{L}=\sum_{S=h^0, A^0}\,\Bigg(
\frac{1}{2}\,\partial_{M}\,S(x,y)\,\partial^{M}\,S(x,y)-
V(S,\phi_L) \,\Bigg) \, ,
\end{eqnarray}
with $M=0,1,...,4$, and
\begin{eqnarray}
\partial_{M}\,S(x,y)\,\partial^{M}\,S(x,y)=\partial_{\mu}\,S(x,y)
\partial^{\mu}\,S(x,y)-\,
(\frac{d S(x,y)}{dy})^2 \, ,\nonumber
\end{eqnarray}
with $\mu=0,1,2,3$. Here we consider the potential $V(S,\phi_L)$,
\begin{eqnarray}
V(S,\phi_L)=\frac{1}{2}\,m_S^2\,
S^2(x,y)+\frac{1}{2}\,g\,S^2(x,y)\,\phi_L^2 \, , \label{Pot1}
\end{eqnarray}
where $\phi_L$ is the localizer singlet field, which is chosen as
$\phi_L=2\,\mu^2\,y$ (see \cite{Hamed}). It has a linear
dependence to the new coordinate $y$ for the points near to the
fixed point and it couples weakly to new scalar fields $S(x,y)$
\footnote{We consider an additional interaction term
$\frac{1}{2}\,g\,\,\phi_2^{\dagger}\,\phi_2\,\phi_L^2$ in the full
lagrangian.}. Now the lagrangian $\emph{L}$ can be rewritten as
\begin{eqnarray}
\emph{L}=\frac{1}{2}\,\partial_{\mu}\,S(x,y)\,\partial^{\mu}\,S(x,y)-
\frac{1}{2}\, S(x,y)\,\Bigg(-\frac{d}{dy^2}+\frac{\partial^2
V}{\partial S^2}\Bigg)\,S(x,y) \, .
\end{eqnarray}
%
%
%
With the expansion over the KK modes as
\begin{eqnarray}
S(x,y)=\sum_r h_r(y)\,S^{(r)}(x)\,,
\end{eqnarray}
one gets the mass equation
\begin{eqnarray}
\Bigg(-\frac{d^2}{dy^2}+m_S^2+ g\,(2\,\mu^2)^2\, y^2
\Bigg)\,h_r(y)=m_r^2\, h_r(y)\,.
\end{eqnarray}
The ground state (the zero mode) solution reads
\begin{eqnarray}
h_0(y)=N\, e^{-\beta\, y^2}\, , \label{f0}
\end{eqnarray}
with the normalization constant $N=\frac{(2\,\beta)^{1/4}}{
\sqrt{\pi^{1/2}\,Erf[\pi\,R\,\sqrt{2\,\beta}]}}$ and its
corresponding mass $m_{0,\,S}$
\begin{eqnarray}
m_{0,\,S}^2=m_S^2+2\,\beta \,, \label{Zer0Modemass}
\end{eqnarray}
with $\beta= \sqrt{g}\,\mu^2$. Here $\mu$ is the energy scale
which adjusts the localization of leptons in the extra dimension,
namely, $\mu=\frac{1}{\sqrt{2}\,\sigma}$.

The KK modes for $n\neq 0$ can be obtained by using the operator
$D^=\frac{d}{dy'}-y'$ as
\begin{eqnarray}
h_r(y)=N^{(r)}\, D^r\,e^{-\frac{1}{2}\, y'^2}\, , \label{fn}
\end{eqnarray}
where $y'=\sqrt{2\,\beta}\,y$ and $N^{(r)}$ is the normalization
constant for $r_{th}$ KK mode, \\
$N^{(r)}=\frac{\beta^{1/4}}{\sqrt{\pi^{1/2}\,2^{
n-\frac{1}{2}}\,(n)!}\,\sqrt{Erf[\pi\,R\,\sqrt{2\,\beta}]}}$.
Finally the masses of the KK modes, including the zero one,
becomes \footnote{Notice that we take even KK modes of new Higgs
bosons, namely the ones with $r=2\,n, \, n=0,1,...$, since the
extra dimension is compactification on the orbifold $S^1/Z_2$. In
this case, the KK mode masses read
$m_{r,\,S}^2=m_S^2+2\,\beta\,(4\,n+1)$ }
\begin{eqnarray}
m_{r,\,S}^2=m_S^2+2\,\beta\,(2\,r+1) \, . \label{KKModemass}
\end{eqnarray}
\newpage
\newpage
\newpage
\end{document}